\def\bdt{\dot \beta}
\def\adt{\dot \alpha}



\def\a{\alpha}
\def\b{\beta}

\def\d{\delta}

\def\f{\phi}
\def\h{\eta}

\def\m{\mu}
\def\n{\nu}

\def\r{\rho}

\def\th{\theta}

\def\beq{\begin{equation}}\def\eeq{\end{equation}}
\def\beqa{\begin{eqnarray}}\def\eeqa{\end{eqnarray}}
\def\barr{\begin{array}}\def\earr{\end{array}}

\def\del{\partial}
\def\Db{\bar{D}}

\parindent=10pt
\baselineskip=18pt
\magnification=1200

\rightline{CERN-TH/96-199}
\rightline{Goteborg-ITP-96-11}
\rightline{King's College-96-14}
\rightline{hep-th/9607239}
\rightline{July 1996}
\vskip .5cm
\centerline {\bf{{Superconformal Ward Identities and $N=2$ Yang-Mills
Theory}}}
\vskip 1cm
\centerline{\bf P.S. Howe ${}^1$}
\vskip .2cm
\centerline{CERN}
\centerline{Geneva, Switzerland}
\vskip .2cm
\centerline{and}
\vskip .2cm
\centerline{\bf P.C. West ${}^1$}
\vskip .2cm
\centerline{Department of Mathematics}
\centerline{King's College, London, UK}
\centerline{and}
\centerline{Institute of Theoretical Physics,
 Chalmers University of Technology,}
\centerline{S-412 96, Goteborg, Sweden.}
\vskip .5cm
\leftline{\sl Abstract}
A reformulation  of the superconformal Ward identities that
combines all the superconformal currents and the associated parameters in one 
multiplet is given for
theories with rigid $N=1$ or $N=2$ supersymmetry. This form of 
the Ward Identities
is applied to spontaneously broken
$N=2$ Yang-Mills theory and used to derive a condition on the
low energy effective action. This condition is satisfied by the
solution proposed by Seiberg and Witten.
\vskip .5cm
\leftline{CERN-TH/96-199, July 1996}
${}^1 $ Permanent address: 
Dept. of Mathematics,
King's College,
London. 
\vfill\eject

\parskip=18pt

{\bf{{0.  Introduction  }}}
\medskip
Some tine ago it was shown that certain four-dimensional supersymmetric
theories are quantum mechanically superconformally invariant or,
equivalently, finite. These theories are: $N=4$ Yang-Mills theory [1],
a large class of
$N=2$ Yang-Mills theories coupled to $N=2$ matter [2]  and certain
$N=1$ theories [3]. It was also shown that in perturbation
theory  the $N=2$ Yang-Mills beta-function has only  one-loop
contributions [2,4]. Although the significance of these results was
not  immediately apparent,  a number of other developments have  taken
place which have focused attention on these theories and which have
relied upon their conformal properties. In particular, the
electro-magnetic duality conjecture of  Montonen and Olive [5] is
believed to be most likely [6] to be valid in these superconformally
invariant theories as the couplings do not run under a change of 
scale and so any symmetry that  inverts the coupling  makes sense at
all scales [7]. It has also been suggested that there may be further 
examples of $N=1$ theories [24] and 
$N=2$ theories [8] which have non-trivial fixed points. 
\par
Recently [9], it has been found that one can determine the low
energy effective action of spontaneously broken $N=2$ Yang-Mills
theory. In practice this means that part of the effective action which
is a chiral sub-integral depending holomorphically on the field
strength superfield, $A$, of the unbroken $U(1)$. Of course, this is
only a part of the full effective action (for the $U(1)$ fields) which
is an integral over all of superspace of a function of $A$ and
$\bar{A}$ and derivatives and which  in general has non-local
contributions. The determination of the low-energy effective action in
[9] makes essential use of properties of the theory related to
electromagnetic duality.
\par
An at first sight unrelated development was the work of BPZ [10],
who  solved a large class of conformal invariant theories in two
dimensions. This represented the first systematic
non-perturbative solution of a class of quantum field theories. The
most likely theories in four dimensions for which one might try to
emulate this achievement are the theories which have extended
supersymmetry. An early signal that such progress may be possible was
the discovery that at a fixed point the anomalous dimension of a
chiral superfield was fixed in terms of its  weight under $R$-symmetry
transformations [11].
\par
Recently [12], the authors have investigated the consequences of
superconformal invariance for certain Green's functions in  rigidly
supersymmetric theories in four spacetime dimensions. It has been
argued [12] that in superconformal theories one can solve for the
Green's functions in  the chiral or  analytic sectors. Chiral sectors
occur in
$N=1$ matter and Yang-Mills theories and in $N=2$ Yang-Mills theories 
and this result generalises the analogous result [13] in two
dimensions.  Analytic sectors occur in  $N=2$ matter (coupled to
Yang-Mills) and $N=4$ Yang-Mills theory. Indeed, in the latter theory
a  large class
of Green's functions are analytic and so the theory is at least
partially soluble.
\par
Although $N=2$ Yang-Mills theory is not superconformally
invariant there are still Ward Identities corresponding to
superconformal transformations which have appropriate
anomalous contributions. In this paper  we derive the anomalous
superconformal
Ward identity and find the conditions  that it implies
for the low energy  effective
action of the spontaneously broken theory. The latter has the form
$\int d^4 x\int d^4 \theta F(A) + c.c $, and we show that 
the superconformal Ward identity
implies that $F$ satisfies
$$ 
a{\del F\over \del a}-2F =8 \pi i\beta_1 u. 
$$
for any gauge group, where $\beta_1$ is the coefficient of the one
loop beta function and $a=<A>$. This constraint is  
indeed satisfied by
the solution for
$F$ found by Seiberg and  Witten. This was first shown for the case
of gauge group $SU(2)$ spontaneously broken to $U(1)$ in reference
[14] and for  the $SU(N)$, $SO(N)$ and $Sp(N)$ gauge groups
spontaneously  broken to
their Cartan subalgebras  in the presence of
certain $N=2$ matter in references [15,16].   In these papers
[14,15,16], the above condition has been observed to hold
phenomologically, that is, the authors have assumed 
 the Seiberg-Witten solution with its associated hypereliptic curve
and  have derived the above  simple condition from it.  In this
paper,  for the first time this condition, and therefore part of the
information about the  low energy effective action, will be   derived
directly from the underlying field theory without  assuming 
electromagnetic duality.
\par
In order to derive this result, we give a new  superspace
formulation of the superconformal currents and
their corresponding Ward identities. It is well known [17] that the
superymmetry and internal symmetry currents and the energy-momentum
tensor belong to the supercurrents $J_{\alpha \dot \alpha}$
and $J$ for $N=1$ and
$N=2$ respectively. In a  superconformal theory, moreover, moments of
the supersymmetry currents and the energy-momentum tensor are 
also conserved. In
section one, we  construct a supermultiplet of moments 
by combining  all the superconformal currents and their
parameters into a single superfield. For $N=1$
theories this superfield is
$f^{\alpha \dot \alpha}J_{\alpha \dot \alpha}$ where
$f^{\alpha \dot \alpha}$ is a superfield that contains the
superconformal parameters. A more complicated expression is given for
the case of $N=2$ supersymmetry.
\par
In sections two and three the form of the superconformal Ward
identities  are derived for  rigid supersymmetric
$N=1$ and $N=2$  theories respectively. This is first done for
the Ward identities that involve the supercurrents
$J_{\alpha \dot \alpha}$ and $J$ and then for the form that
involves all the superconformal currents and the associated 
parameters. The advantage of this latter
formulation is that it allows one to deal with all the superconformal
symmetries together in a simple way.
\par
In section four, we apply the superconformal Ward identities to the
spontaneously broken $N=2$ Yang-Mills theory and derive a superfield 
constraint on the Seiberg-Witten solution. We find that one special
case  of this equation is the constraint equation 
given above. We then derive an alternative form of this equation by
replacing the anomaly part by a derivative with respect to a conformal
compensating field.  
\medskip
{\bf{{1. Currents in Supersymmetric Theories}}}
\medskip
In this section we review the currents that appear in $N=1$  and $2$
supersymmetric theories. We begin by considering a current
corresponding to an internal symmetry. In $N=1$ supersymmetric
theories such an internal symmetry  current belongs to a real linear
supermultiplet $L$ and the conservation condition is $\Db^2L=0$.  For
an
$N=2$ supersymmetric theory, on the other hand, an internal symmetry
current is a component of a real symmetric superfield,  $L_{ij}$,
which transforms under the triplet representation of $SU(2)$.  The
current conservation condition is
$D_{\alpha(i} L_{jk)}=0$.  For both of these multiplets the  internal
(spacetime) currents occur at the $\theta \bar \theta $ level and  so
the superfields $L$ and $L_{ij}$ have dimension 2. We observe that an
internal symmetry current has superpartners that have no 
interpretation as conserved currents in spacetime.

Let us now consider the supersymmetry currents,   $\{j_{\alpha i}\},\
i=1,\dots,N$, and the supermultiplets they belong to. The
supersymmetry variation of 
$j_{\alpha i}$ has the form
$\delta j_{\alpha i}\sim [j_{\alpha i},  \epsilon^{\beta j}Q_{\beta
j}]$ and must, when integrated, give the correct relation for the
anti-commutators of two supersymmetry charges that occur  in the
supersymmetry algebra. We therefore conclude that the supersymmetry
current is in the same multiplet as the energy-momentum tensor [17]. 
By extending this argument we find the supercurrent multiplet
corresponding to superconformal symmetry also includes
the internal symmetry currents whose charges appear in the
supersymmetry algebra. For $N=1$ there is only one such current,  the
$R$ current, while for $N=2$ there is the $R$ current as well as the
currents corresponding to the internal $SU(2)$ symmetry. Proceeding in
this manner one can systematically construct the supercurrent
multiplet associated with superconformal symmetry in a purely
algebraic fashion [30]. In addition
 to the conserved currents listed above one also finds that the 
components of the supercurrent include certain moments of the
energy-momentum tensor and of the supersymmetry currents. 

The $N=1$ supercurrent is described by a real superfield
$J_{\alpha \dot \alpha}$. It has
dimension 3 and obeys, in the absence of  any
anomalies, the conservation condition
$$
\Db^{\dot\alpha}J_{\alpha\dot\alpha}=0. \eqno(1.1)
$$
For $N=2$ the supersymmetry current belongs to a real scalar 
superfield $J$ which has dimension 2 and which, again in the absence
of anomalies, obeys the equation
$$D_{ij}J=0. 
\eqno (1.2)
$$
where
$$
D_{ij}:=D_{\a i}D^{\a}_j .
\eqno(1.3)
$$
\par
For both
$N=1$ and $N=2$ superconformal theories, the parameters  associated
with superconformal transformations can be combined in a superfield of
parameters,
$f^{\alpha\dot\alpha}$, which is subject to the constraint
$$
D_{(\alpha i}f_{\beta)\dot\beta}=0.
\eqno(1.4)
$$
Such a parameter determines a (real) superconformal  Killing vector
field $X_f$ given by
$$
X_f=f^{\a\adt}\del_{\a\adt}+\f^{\a i}\del_{\a i} -\bar\f^{\adt}_i
\Db_{\adt}^i
\eqno(1.5)
$$
where
$$
\f^{\a i}:=-{i\over2}\Db_{\adt i} f^{\a\adt}.
\eqno(1.6)
$$
By definition, a superconformal Killing vector field is one  which
preserves chirality in super Minkowski space, i.e. one which satisfies
$$
[\Db_{\adt i},X_f]\sim \Db_{\adt i}.
\eqno(1.7)
$$

The  components of $f^{\alpha\dot\alpha}$ can be defined to be   given
by the following superfields evaluated at $\theta=0$:
$$
\eqalign{
v^{\alpha\dot\alpha}&:=f^{\alpha\dot\alpha}\vert ,\cr
\bar \eta^{\dot\alpha}_i&:
={i\over2}D_{\alpha i}f^{\alpha\dot\alpha}\vert ,\cr
\eta^{\alpha i}&:
=-{i\over2}\Db_{\dot\alpha}^i f^{\alpha\dot\alpha}\vert , \cr
w_{ij}&:=[D_{\alpha i},\Db_{\dot\alpha j}]
f^{\alpha\dot\alpha}\vert,\cr}
\eqno(1.8)
$$
where the vertical bar denotes evaluation of a superfield at $\th=0$.
It follows from equation (1.3) that the fields
$v^{\alpha\dot\alpha}$,  $\eta^{\alpha i}$ and its complex
conjugate  obey the conformal Killing equation and the spinor Killing
equation respectively.  That is,
$$
\del_{(\a(\adt} v_{\b)\bdt)}=0
\eqno(1.9)
$$
and
$$
\del_{(\a |\adt|}\h_{\b)}^i=0.
\eqno(1.10)
$$
Solving equations (1.9) and (1.10),  one sees that
$v^{\alpha\dot\alpha}$ contains translations, dilations and
special conformal transformations, while $\eta^{\alpha i}$
and its complex conjugate contain ordinary ($Q$) and special  ($S$)
supersymmetry transformations. The $x$-independent field
$w_{ij}$ is the parameter of internal symmetry transformations of the
supersymmetry algebra.

In a conformal field theory, the energy-momentum  tensor $T_{\m\n}$,
assumed to be symmetric, traceless and conserved, can be combined with
the parameter of conformal transformations, $f^{\m}$, into a conserved
current $J_{\m}^f$, $J_{\m}^f:=f^{\n}T_{\m\n}$, the conservation of
$J_{\m}^f$ being a consequence of the constraints 
on $T_{\m\n}$ and the
fact that $f^{\m}$ is a conformal Killing vector,
$$
\del_{(\m} f_{\n)}={2\over d}\h_{\m\n}\del_{\r}f^{\r},
\eqno(1.11)
$$
where $d$ is the dimension of spacetime.
Similarly, in an $N=1$ superconformal field theory,
the  currents and their associated parameters can be neatly packaged
into a real superfield
$J^f=f^{\alpha\dot\alpha}J_{\alpha\dot\alpha}$.
We observe that if $J_{\alpha\dot\alpha}$ is not anomalous then  $J^f$ is
a linear multiplet, i.e.
$$
\bar{D}^2J^f=0.
\eqno (1.12)
$$
We note that the superconformal currents and  their parameters when
packaged in $J^f$ have the same dimension and  obey the same
conservation equation  as an internal symmetry current $L$ which was
discussed at the beginning of the section.

In an $N=2$ superconformal theory we may likewise combine all the
superconformal currents and their
associated parameters into a real  superfield $J^f_{ij}$
which is symmetric in its $SU(2)$ indices $i,j$.  This superfield is 
$$
J^f_{ij}= i\left(f^{\alpha\dot\alpha}D_{\alpha i}\bar{D}_{\alpha
j}J+{1\over2}(D _{\alpha i}f^{\alpha\dot\alpha})\bar{D}_{\dot\alpha
j}J-{1\over2}(\bar{D}_{\dot
\alpha i}f^{\alpha\dot\alpha})D_{\alpha j}J+{1\over4}(D_{\alpha
i}\bar{D}_{\dot
\alpha j}f^{\alpha\dot\alpha})J\right)
\eqno (1.13)
$$
The coefficients in the above expression are determined by  the
requirement that
$$
D_{\alpha (i}J^f_{jk)}=0,
\eqno(1.14)
$$
if the supercurrent $J$ is  conserved, that is, satisfies
equation (1.2). Thus the $N=2$ superconformal currents follow the
same pattern as  the
$N=1$  superconformal currents; namely the superconformal currents and
their parameters combine into a superfield which is of the same type
as an internal symmetry current.
\par
In this paper,
we want to consider  theories which are classically  superconformal
invariant, but which develop anomalies quantum mechanically.   The
theories of interest to us  are invariant under Poincar\'{e} and
$Q$-supersymmetry transformations, but have anomalies in some, or all,
of the remaining superconformal symmetries. In the presence of
anomalies the supercurrent no longer obeys equation (1.1) or (1.2) 
which become modified. Although there are several possible types of 
anomaly for
$N=1$ supersymmetric theories, we will assume that the supercurrent
$J_{\alpha\dot \alpha}$ obeys the operator equation
$$
\Db^{\dot\alpha}J_{\alpha\dot\alpha}=D_\alpha S
\eqno(1.15)
$$
where $S$ is a chiral superfield $(\Db_{\dot\alpha}S=0)$.
For $N=2 $ supersymmetric Yang-Mills theories, we will show that
the anomaly  is of the form
$$
D_{ij}J=\bar{D}_{ij}\bar{S}
\eqno(1.16)
$$
where $S$ is a chiral superfield, $\Db_{\adt i}S=0$.
\par
Anomalies in the supercurrent modify the conservation
equations (1.12) and (1.13). It is straightforward to verify that if
$J_{\alpha\dot\alpha}$ has the conformal anomaly of equation (1.15)
then
$$
\bar{D}^2J^f=-\bar{D}_{\dot\alpha}f^{\a\dot\alpha}
D_\a S+if^{\alpha\dot
\alpha}\partial_{\alpha\dot\alpha}S,
\eqno(1.17)
$$
where ${D}^2={1\over 2}{D}_{\alpha} D^{\alpha},\
\bar{D}^2=-{1\over 2}\bar{D}_{\dot\alpha}\bar{D}^{\dot\alpha}$. 
The corresponding
equation for $N=2$ implies that
$J_{ij}$ obeys the equation
$$
D_{\alpha (i}J^f_{jk)}
=-{i\over4}\bar{D}_{\adt(i}f_\alpha{}^{\adt}
D_{jk)}J=-{i\over4}\bar{D}_{(ij|}(\bar{D}_{\adt
|k)} f_\alpha{}^{\adt} \bar{S}).
\eqno(1.18)
$$
\vfill\eject
\medskip
{\bf{{2. \ N=1 Supersymmetric Ward Identities}}}
\medskip
Let us consider an $N=1$ supersymmetric theory which  contains  a set
of   chiral superfields collectively denoted by
$\varphi$, their conjugates $\bar{\varphi}$ and the  Yang-Mills
potential
$V$. The most general renormalizable action is of the form
$$ \int d^4xd^4\theta\,\bar{\varphi}e^{gV}\varphi
+ Im \{ {\tau\over 4\pi}  \int d^4xd^2\theta\   {\rm tr}(W_\alpha
W^\alpha) \}
 +\int d^4xd^2\theta\, U(\varphi)+{\rm{c.c.}}.
\eqno(2.1)$$
where $U$ is the superpotential  which is at most cubic in the chiral
superfields and $\tau = {\theta \over 2\pi} + i{4\pi\over g^2}$.
\par 
We denote the effective action of this theory by
$\Gamma$. We wish to consider the constraints placed on
$\Gamma$ by superconformal
symmetry, that is, by the superconformal Ward identity.  This
subject has been studied extensively in the past, see for example
references [18,19] and references therein. Here we will give a simple 
derivation of the WI based on a superspace version of Noether's
identification of the current. We shall ignore the complications which
arise due to gauge-fixing and ghosts as these will play no part in the
applications in this paper.

We therefore consider the
variation of $\Gamma$ under the superspace analogue   of
(infinitesimal) local reparametrizations.
The parameter of these transformations is a spinor  superfield
$L^\alpha$. The chiral superfield $\varphi$, which obeys a flat
space chiral condition and has $R$ weight $q$,
transforms under
$L^\alpha$ transformations as
$$
\eqalign{
\delta\varphi &=-\bar{D}^{\dot\alpha}
L^\alpha\partial_{\alpha\dot\alpha}\varphi
-i\bar{D}^2L^\alpha D_{\alpha}\varphi
+q\triangle\varphi \cr
&=-i\bar{D}^2(L^\alpha D_\alpha\varphi -qD_\alpha
L^\alpha\varphi),\cr}
\eqno(2.2)$$
where
$$\triangle
=-\partial_{\alpha\dot\alpha}\Db^{\dot\alpha}L^\alpha
+iD_\alpha \bar{D}^2 L^{\a}.
\eqno(2.3)$$
The Yang-Mills potential transforms as
$$
\eqalign{
\delta V &=-i(\bar{D}^2 L^\alpha)D_\alpha
V-{1\over2}\bar{D}^{\dot\alpha}L^\alpha
\partial_{\dot \alpha \alpha}V \cr
&= -i\bar{D}^2 (L^\alpha D_\alpha V)
+{i\over2}\bar{D}^{\dot\alpha}L^\alpha [D_
\alpha,D_{\dot\alpha}] V+iL^\alpha W_\alpha+{\rm{c.c.}}\cr }
\eqno(2.4)
$$
By definition, the supercurrent $J_{\alpha\dot\alpha}$
couples to the
supergravity superfield $H^{\alpha\dot\alpha}$ at linearised order in
$H^{\alpha\dot\alpha}$ in the form
$$2\int d^4xd^4\theta\
H^{\alpha\dot\alpha}J_{\alpha\dot\alpha}
\eqno(2.5)$$
Since the
transformation of $H^{\alpha\dot\alpha}$ is given by
$$\delta H^{\alpha\dot\alpha}
=-{i\over2}(D^\alpha \bar{L}^{\dot\alpha}+\Db^{\dot\alpha}
L^\alpha)
\eqno(2.6)$$
the variation of (2.5) contributes
$$-i\int d^4xd^4\theta\ \bar D^{\dot\alpha}L^\alpha
J_{\alpha\dot\alpha} +{\rm{c.c.}}
\eqno(2.7)$$
to zeroth order in $H^{\alpha\dot\alpha}$.
This variation must be cancelled by the variation of  $\Gamma$ under
the transformation of equations (2.2) and (2.4) and we will  regard
this variation as  the definition of the supercurrent associated with
$\Gamma$. To be precise, we take the supercurrent to be defined by
$$\delta\Gamma
=i\int d^4xd^4\theta\ (\Db^{\dot\alpha}L^\alpha )J_{
\alpha\dot\alpha} +{\rm{c.c.}}
\eqno(2.8)$$

We now find an expression for $J_{\alpha\dot\alpha}$
using the method explained above. From equations (2.2) and (2.4) we
find that
$$\eqalignno{
\delta\Gamma&
=\int d^4xd^4\theta\bigg\{\{L^\alpha D_\alpha\varphi-qD_\alpha L
^\alpha\varphi\}{{\delta\Gamma}\over{\delta V}}\cr
& +(-i\bar{D}^2(L^\alpha D_\alpha V)
+{i\over2}\bar{D}^{\dot\alpha}
L^\alpha[D_\alpha,D_{\dot\alpha}]V
+iL^\alpha W_\alpha ){{\delta\Gamma}\over{
\delta V}}\bigg\} + c.c &(2.9)}$$
where $q$ is the $R$ weight of $\varphi$ which must be
$1\over3$ if we have
one chiral superfield with a cubic superpotential. If we restrict 
our attention to a
$U(1)$ gauge theory then gauge invariance implies that
$$\bar{D}^2{{\delta\Gamma}\over{\delta V}}=0
\eqno(2.10)$$
and so we can discard the first term in the second bracket above. In
what follows we will consider the Abelian theory, but the
modification to the non-Abelian case can be made.

Having identified the supercurrent $J_{\alpha\dot\alpha}$ in terms of
the variations, it is straightforward to write down the Ward identity.
For a superconformal theory, there are no
anomalies and the Ward identity is given by setting the
$\delta\Gamma$ of  equation (2.8) equal to that of equation (2.9).
For theories with superconformal anomalies the
supercurrent $J_{\alpha\dot
\alpha}$ will obey the operator equation (1.15) which
 is valid in Green's functions. In recovering the
Ward identity from this operator equation we find additional terms
which arise due to  the fact that the  derivatives are outside  the
time ordering of the Green's function  so that they  act on the time
ordering as well as the current. These extra (contact) terms are none 
other than the variations of the fields, i.e. the $\delta\Gamma$ of
equation (2.9).

Taking the anomaly into account we therefore find that the
Ward
Identity for an $N=1$ rigid supersymmetric theory for the above transformations is 
given by
$$\eqalignno{\int d^4xd^4\theta
\big\{(L^\alpha D_\alpha\varphi-qD_\alpha L^
\alpha\varphi){{\delta\Gamma}\over{\delta\varphi}}\big\}
&+(iL^\alpha W_\alpha){{\delta\Gamma}\over{\delta V}}
-i(\Db^{\dot\alpha} L^\alpha J_{\alpha\dot\alpha
})\cdot\Gamma+{\rm{c.c.}}\cr
&=+i\int d^4xd^4\theta(L^\alpha D_\alpha S)\cdot\Gamma+{\rm{c.c.}}
&(2.11)}
$$
\par
In this equation we have redefined the supercurrent by
$$J_{\alpha\dot\alpha}\rightarrow
 J_{\alpha\dot\alpha}-{1\over2}[D_\alpha,D_{
\dot\alpha}]V{{\delta\Gamma}\over{\delta V}},\eqno(2.12)$$
in order to obtain a supercurrent which is gauge invariant.
The non-gauge
invariance of the supercurrent is an artifact of the way we have
derived it. Essentially, we have used the coupling of the theory to
supergravity so that the superfields and their variations
can involve spinorial covariant derivatives that contain the supergravity
superfield
$H^{\alpha \dot \alpha}$. We have chosen the chiral superfield to
obey a flat-space chirality condition; however,
 the Abelian gauge 
invariance is realised with a chiral parameter
$\Lambda$ that satisfies
${\bf {\Db}}_{\dot\alpha}\Lambda=0$ where $\bf {\bar D}_{\dot\alpha}$ 
is the spinorial covariant derivative which involves $H^{\alpha \dot
\alpha}$. The flat space chiral parameter
$\Lambda_0$, satisfying
$\Db_{\dot\alpha}\Lambda_0=0$, is related to $\Lambda$ by
$$\Lambda=\Lambda_0
-{i\over2}H^{\alpha\dot\alpha}\partial_{\alpha\dot\alpha}
\Lambda
_0+\dots.
\eqno(2.13)$$
Gauge invariance of the effective action plus the
lowest order supergravity
coupling of equation (2.5) implies that
$$\delta J_{\alpha\dot\alpha}
={i\over2}\partial_{\alpha\dot\alpha}\Lambda_0{{\delta
\Gamma}\over{\delta V}}+{\rm{c.c.}}
\eqno(2.14)$$
>From this equation we can deduce the 
change (2.12) in
$J_{\alpha\dot\alpha}$ required to obtain a gauge-invariant current.

>From the Ward identity of equation (2.11) we can deduce
the unintegrated superconformal WI by
functionally differentiating with respect to $L^{\a}$. It is:
$$D_\alpha\varphi{{\delta\Gamma}\over{\delta\varphi}}
-qD_\alpha\left(\varphi{{
\delta\Gamma}\over{\delta\varphi}}\right)
+iW_\alpha{{\delta\Gamma}\over{\delta
V}}-i\bar{D}^{\dot\alpha}J_{\alpha\dot\alpha}\cdot\Gamma
=-iD_\alpha S\cdot\Gamma
\eqno(2.15)$$

We can derive this Ward identity by the following argument. 
Given that the identity contains only gauge invariant quantities it
must involve only
$\varphi, \bar \varphi,W_\alpha$ and $S$ and $\Gamma$. The Ward
identity consists of three  types of term: terms that correspond  to
the variation of the fields, a term of the form
$\bar{D}^{\dot\alpha}J_{\alpha\dot\alpha}\cdot\Gamma$ and a
term with the anomaly $S$. The last two terms must be such that the
anomaly equation (1.15) holds in Green's functions. This leaves
only the first type of
term which we can fix by using dimensional analysis and by demanding
that the coefficients agree with those of the free theory.

We now give an alternative form of the Ward identity that
includes the parameters
of the superconformal transformations and involves the current $J^f$ 
defined in section one. 
The advantage of this formulation is that one can make direct contact
with the variation of the effective action under superconformal
transformations and one can include all transformations in one Ward
Identity.
The transformation of $\varphi$ and $V$ under superconformal
transformations are given by
$$
\eqalign{
\delta \varphi
&=X_f\varphi +q\triangle \varphi,\cr
\delta V &=X_f V,}
\eqno(2.17)
$$
where $X_f$ is a superconformal Killing vector  (note that it
simplifies when acting on chiral fields) and
$$ 
\triangle=
\partial_{\alpha \dot \alpha}f^{\alpha\dot \alpha}-D_\alpha \f^\alpha.
\eqno(2.18)$$

We can deduce the required form of the Ward identity by
substituting
$$L^\alpha=f^{\alpha\dot\gamma}\bar{D}_{\dot\gamma}\delta^8(z-z^\prime)
+{1\over2}
\bar{D}_{\dot\gamma}f^{\alpha\dot\gamma}\delta^8(z-z^\prime)
\eqno(2.19)$$
into equation (2.11) to get
$$\eqalignno{
\delta\varphi{{\delta\Gamma}\over{\delta\varphi}}&
+\left(f^{\alpha\dot\gamma}W_
\alpha \Db_{\dot\gamma}-{1\over2}\Db_{\dot\gamma}
f^{\alpha\dot\gamma}W_\alpha\right
){{\delta\Gamma}\over{\delta V}}-\bar{D}^2J^f
=\bar{D}^{\dot\alpha}f^\gamma_{\dot
\alpha}D_\gamma S\cr
&-if^{\alpha\dot\alpha}\partial_{\alpha\dot\alpha}S
+\left[{1\over3}\partial_
{\alpha\dot\alpha}
\left(f^{\alpha\dot\alpha}\varphi{{\delta\Gamma}\over{\delta
\varphi}}\right)+{i\over6}D_\alpha
\left(\Db_{\dot\alpha}f^{\alpha\dot\alpha}
\varphi{{\delta\Gamma}\over{\delta\varphi}}\right)\right],&
(2.20)}$$
where $\delta\varphi$ is given in equation (2.16). We note that the
final term  in brackets on the right hand side is a total derivative.

Finally, we can find the integrated form of this equation by
integrating over chiral superspace and adding the complex conjugate. 
The result is 
$$\int d^4xd^2\theta
\left\{\delta\varphi{{\delta\Gamma}\over{\delta\varphi}}+
\left(f^{\alpha\dot\gamma}W_\alpha \Db_{\dot\gamma}
-{1\over2}\Db_{\dot\gamma}f^{
\alpha\dot\gamma}W_\alpha\right) {{\delta\Gamma}\over{\delta
V}}\right\}+{\rm{ c.c.}}=2i\int d^4xd^2\theta\triangle S\cdot\Gamma
+{\rm{c.c.}}\eqno(2.21)$$
where $\triangle$ is defined in  equation (2.18). 
The current term
does not contribute since $\ \ \ \ \ \ \ \ \ $ $ [D^2,\bar{D}^2]J^f $
is a total space-time derivative.
\par
An alternative form of the anomalous Ward identity [18,19] 
can be given
by including, in addition to the supergravity field
$H^{\alpha \dot
\alpha}$,  the chiral compensator $\phi$. In this case,
equation (2.5) generalises to 
$$2\int d^4xd^4\theta\
H^{\alpha\dot\alpha}J_{\alpha\dot\alpha} +
\{ 2\int d^4xd^2\theta \phi S + c.c \}. 
\eqno(2.22)$$
The action of equation (2.1) plus the above term is invariant to 
zeroth order in the supergravity fields if we take  $H^{\alpha \dot
\alpha}$  to vary according to equation  (2.6) and we take  the
variation of
$\phi$ to be given by 
$$ \delta \phi = {i\over2} \bar D^2 D_\alpha L^\alpha.
\eqno(2.23)$$
Clearly, when the matter fields satisfy their equations of motion 
then the variation of the action implies that the supercurrent 
satisfies equation (1.15) from which we recognise $S$ as the 
part of the effective action which is not superconformally invariant,
that is the anomaly. 
\par
Upon varying the effective action plus equation (2.22) we can read
off,  as before, the  coefficient of $L^\alpha$  to find the 
Ward identity
with the anomaly automatically encoded. As such, we recover 
 equations (2.15), (2.20) and (2.21). 
\par
The chiral compensator $\phi$  contains the component fields (4+4)
 required to complete the conformal
supergravity multiplet of fields (8+8) to the old minimal
Poincar\'e supergravity theory (12+12).  As such, it  necessarily
couples to the chiral anomaly as shown  in equation (2.22).  
\par
Clearly, we can keep the dependence of the effective action on 
the supergravity fields  $H^{\alpha \dot \alpha}$ 
and  $\phi$ and replace the presence of the current and the anomaly 
in the Ward identity 
by suitable functional derivatives with respect to the
supergravity fields  and then set the supergravity fields  to zero. 
In particular, one can carry out this procedure
for the anomaly alone. 
\medskip
{\bf{{3. \ Ward Identities for N=2 Supersymmetric Yang-Mills
Theory}}}
\medskip
The $N=2$ supersymmetric Yang-Mills theory [20] is described by a
complex scalar superfield
$W$ which transforms under the adjoint representation of the gauge 
group. This superfield is covariantly chiral, i.e.
$\bar{\nabla}_{\dot\alpha}^i W=0$, and satisfies the constraint
$$ \nabla_{ij} W=\bar{\nabla}_{ij}\bar{W}
\eqno(3.1 )$$
where
$$
\nabla_{ij}=\nabla_{\a (i}\nabla^{\a}_{j)},
\eqno(3.2)
$$
$\nabla_{\a i}$ is the spinorial covariant derivative including 
the gauge connection. The components of the superfield $W$  are a 
complex scalar field, spinor fields in an $SU(2)$  doublet, the 
field strength tensor of the spacetime gauge field, and an $SU(2)$
triplet of auxiliary fields. We shall denote the superspace field
strength tensor in the Abelian case by $A$.

The constraints (3.1) are solved in the Abelian case by
$$
A=\bar{D}^4D^{ij}V_{ij},
\eqno(3.3)
$$
where $V_{ij}=V_{ji}$ is the (unconstrained) superfield  prepotential
that contains the spacetime gauge
potential [21]. The solution in the non-Abelian case is more
complicated, but can still be written in terms of an unconstrained
superfield $V_{ij}$ [22]. Alternatively, one can use the harmonic
superspace formalism [23] which allows one to use a prepotential of
dimension zero, but we shall not consider this possibility here.

For the free theory, the action is given by
$$ \int d^4xd^4\theta A^2+{\rm{c.c.}},\eqno(3.4)$$
and the supercurrent is given by
$$J=A\bar{A}\eqno(3.5)$$
It is easy to check that it is conserved, i.e.
$D_{ij}J=0$, by virtue of the equation of motion $D_{ij}A=0$.

We now derive the Ward identity for the $N=2$ supersymmetric 
Yang-Mills theory. We could do this, as for the $N=1$ case, by
considering the variation of the action under super
reparametrisations. However, the structure of $N=2$ superspace
supergravity is  significantly more complicated than that of
$N=1$ supergravity and we shall not give the details of this 
approach here. Instead, we shall derive the identity heuristically
using gauge invariance and dimensional analysis as we did for the
$N=1$ case. The Ward identity  must again have the
same three types of term and the ones involving the current and the
anomaly must be consistent with
the relation between the current
$J$ and the anomaly
$\bar{S}$ of equation (1.16) in the sense that this latter equation is
realised  as an operator equation in Green's functions. The first
type of terms
which are associated with   the variation of the effective action
must contain $\displaystyle{{{\delta\Gamma}
\over{\delta V^{ij}}}}$, a function of $A$ and possible
covariant derivatives
acting on these. We note that
$\displaystyle{{{\delta\Gamma}\over{\delta V^{ij}}}}$ has dimension
two, but the dimension of all terms in the Ward identity is the
same as that of the current term which is given by
$D^{ij}J$ and has dimension 3. Taking all this information
into account and fixing the one unknown  coefficient by inserting the
known current for the free classical theory we find that the Ward
 identity is given by
$$\bar{A}{{\delta\Gamma}\over{\delta V^{ij}}}-D^{ij}J\cdot\Gamma
=-\bar{D}^{ij}
\bar{S}\cdot\Gamma
\eqno(3.6)$$
Using the expression for $J_{ij}$ of equation (1.13) we find the Ward
Identity  which corresponds contains all the  superconformal
currents and their parameters; it is given by
$$-{i\over4}\Db_{\dot\gamma(i}f_\alpha{}^{\dot\gamma}
\bar{A}{{\delta\Gamma}\over{ \delta
V^{jk)}}}+{D}_{\alpha(i}J^f_{jk)}\cdot\Gamma
 =-{i\over4}\bar{D}_{(jk}(
\bar{D}_{\dot\gamma i)}f_\alpha{}^{\dot\gamma}\bar{S})\cdot\Gamma.
\eqno (3.7)$$

The effective action for the U(1) field will contain two types of term, one of 
which is a
full integral over superspace and is a function of $A$ and $\bar{A}$
and the other which is an integral only over a chiral sub-integral of
superspace and is a function of $A$ only. We can write the latter
contribution in the form
$$\Gamma_c=\int d^4xd^4\theta F(A)+{\rm{c.c.}}.
\eqno(3.8)$$
The above definition of $F$ differs from that in the literature. To
recover the usual definition we should take $F \to -{i\over 16\pi}
F$.  The absence of the usual factors simplifies all the following 
equations.  We
now focus on the constraints imposed by the Ward  identity on this
latter contribution since this is the  low energy effective action that
appears in the Seiberg-Witten formalism. The variation of
$\Gamma_c$ with  respect to $V^{ij}$ is given by
$${{\delta\Gamma_c}\over{\delta V^{ij}}}
=\bar{D}^{ij}\bar{F}^\prime+ D^{ij}F^\prime,
\eqno(3.9)$$
where $\displaystyle {F^\prime={{\del F}\over{\del A}}}$.

Using the identity
$$\bar{A}\bar{D}_{jk}\bar{F}^\prime
=\bar{D}_{jk}(\bar{A}\bar{F}^\prime-2\bar{F})+D_{jk}
(A\bar{F}^\prime)
\eqno(3.10)$$
we find
that we can write the first term in the Ward identity of equation
(3.6) when restricted to $\Gamma_c$ as
$$\bar{A}{{\delta\Gamma_c}\over{\delta
V^{ij}}}=\bar{D}^{ij}(\bar{A}\bar{F}^\prime-2
\bar{F})+D^{ij}(A\bar{F}^\prime+\bar{A}F')\eqno(3.11)$$
Substituting this expression into equation (3.6) we find the Ward
Identity can be written as
$$\bar{D}^{ij}(\bar{S}+\bar{A}\bar{F}^\prime-2\bar{F})=-D^{ij}
(J-A\bar{F}^\prime-\bar{A}
F')\eqno(3.12)$$
We may rewrite this equation in the form
$$\bar{D}^{ij}\hat{J} =D^{ij}\hat{\bar{S}}
\eqno(3.13)$$
where
$$\hat{J}=J-A\bar{F}^\prime-\bar{A}F',\ \
{\rm {and}}\ \
\hat{\bar{S}}=\bar{S}+\bar{A}\bar{F}^\prime-2\bar{F}
\eqno(3.14)$$

We can regard $\hat{S}$ and $\hat{J}$ as a redefined anomaly
and current respectively. This remarkable simplification of the Ward
identity associated with the restricted  effective action of 
equation (3.8) is
essential for the derivation of the identity which is the subject of
this paper. We now define a corresponding 
$\hat{J}^f_{ij}$ which is given by equation (1.13) except that we
replace $J$ with
$\hat{J}$ so that
$$D_{\alpha(i}\hat{J}^f_{jk)}=-{i\over4}\bar{D}_{\dot\gamma (i}
f_\alpha^{{}\dot\gamma}
D_{jk)}\hat{J}
\eqno(3.15)$$
Using equation (1.4), we find that
$$
{D}_{\alpha
(i}\hat{J}^f_{jk)} =-{i\over4}(\bar{D}_{\dot\gamma(i}f_\alpha{}^{\dot
\gamma})\bar{D}_{jk)}\hat{\bar{S}}
=-{i\over4}\bar{D}_{(jk}(\Db_{\dot\gamma i)}
f_\alpha^{\dot\gamma}\hat{\bar{S}})
\eqno(3.16)$$

To obtain the integrated Ward identity we act with $\int
d^4x\bar{D}^{ij}D^{\alpha k}$ on equation (3.16) and add the complex
conjugate. The term involving
$\hat{J}^f_{ij}$ does not contribute as it is a space-time derivative. 
This leaves
$$
\eqalign{
&\int d^4x\bar{D}^{ij}D^{\alpha k}\bar{D}_{(jk}
(\bar{D}_{\dot\gamma
i)}f_\alpha{}^{\dot\gamma}\hat{\bar{S}})+{\rm{c.c.}}\cr &=\int
d^4x\bar{D}^{ij}\bar{D}_{(jk}
\big[-i\d^k_{i)}\partial_{\alpha\dot\gamma}f^{\alpha\gamma}
+\bar{D}_{\dot\gamma i)}D_\alpha^
kf^{\alpha\dot\gamma}\big ]\hat{\bar{S}}+{\rm{c.c.}}\cr
&={4i\over3}\int
d^4xd^4\bar{\theta}\{\bar{\triangle}\hat{\bar{S}}\}+{\rm{c.c.}}=0.\cr }
\eqno(3.17)$$
In the last step we have used the identities
$$ {\bar{D}}^{ij}{\bar{D}}_{kl}
={1\over 6}(\delta^i_k\delta^j_l
+\delta^j_k\delta^i_l)
\bar{D}^{mn}\bar{D}_{mn},\qquad
\bar{D}_{\dot\alpha(k}\bar{D}_{ij)}=0,
\eqno (3.18)$$
and the definitions
$$\int d^4\bar{\theta}=\bar{D}^{ij}\bar{D}_{ij},\ \
\bar{\triangle}=\partial_{\alpha\dot\alpha}f^{\alpha\dot\alpha}
+{\bar{D}}_{\dot \gamma i}\bar\f^{\dot\gamma i}.
\eqno(3.19)$$
Rewriting (3.17) in terms of S using equation (3.14) and taking the
complex conjugate, we find the final result
$$
\int d^4x d^4\th\, \triangle(A{\del F\over\del A}-2F +S)+{\rm{c.c}}=0
\eqno(3.20)
$$ 
\par
Later in the paper it will also be useful to consider 
   the $N=2$  analogue of   introducing the
supergravity fields discussed for the case of
$N=1$ at the end of the last section. Of
most significance to us will be the r\^{o}le of the supergravity
compensator. The (24+24) set of fields of $N=2$
superconformal supergravity can  be compensated in a number of ways 
to form a (40+40) set of Poincar\'e supergravity fields. However, 
in this procedure one always adds a (8+8) chiral compensator 
which can be represented by a reduced chiral superfield $\phi_r$
subject to  the constraints $\bar D_{\dot \alpha i} \phi_r =0,
\ D^{ij}\phi_r =\bar
D^{ij}
\phi_r$. For a review of this compensation mechanism  we refer the
reader to  reference [29]. 
\medskip
{\bf{4.\  Application of the Ward Identity to Spontaneously
Broken   Yang-Mills Theory}}
\medskip
The authors of reference [9]  considered $N=2\ SU(2)$  Yang-Mills 
theory spontaneously broken to $U(1)$ and, assuming this theory to
exhibit electromagnetic duality, were able  to derive an expression
for the low energy effective action. One way of defining their
low energy effective action would be to regard it as that obtained by
simply carrying out the functional integral for all the massive
fields, but other definitions have been suggested. Whichever method
is considered the low energy effective action is taken to be of
the form of equation (3.8) and so depends holomorphically on only on
the
$N=2$ Abelian superfield $A$.

In the discussions of reference [9] two possible variables are
discussed for the formulation of the above effective action. The
variable $A$ is, as equation (3.8) makes clear, the one in  terms of
which we can write the effective action in a manifestly
$N=2$ supersymmetric manner. The other variable $U$ is defined  by
$U= {1\over 2}{\rm tr}\,W^2$. For the free and perturbative theory the
relation between $a=\langle A\rangle$ and 
$u=\langle U \rangle = {1\over 2} \langle {\rm tr}\,W^2 \rangle$ is simply
$u={1\over 2}a^2$.  However, for the full non-perturbative theory
the relation between
$a$ and
$u$ is complicated. The variable
$a$ viewed as a function of $u$ contains singularities at isolated
points. The determination of the monodromies around these
singularities provides the mechanism [9] for determining
$a$ and $\displaystyle{{{\partial F}\over{\partial a}}}$ as  a
function of $u$  and hence $F$ as a function of $a$. 
\par
When discussing the $N=2$ superconformal transformations  of the
effective action we must realise the transformations in terms of the
$N=2$ superfield $A$ since this is the variable which carries the
standard representation of $N=2$ supersymmetry. As a result, the
superconformal Ward identity is given by equations (3.7) or (3.13)
with the variable $A$ as indicated.

The anomaly ${S}$ must be single valued
with respect to modular transformations
around the singular points and must be a gauge-invariant object. 
Further, the anomaly
${S}$ has dimension two and is a gauge-invariant chiral
superfield. The only possible candidate is
$${S}= {c\over 2} {\rm tr}\,W^2= c U, 
\eqno(4.1)$$
where $c$ is a constant. One can verify that the
above ${S}$ is consistent
with the $R$ transformations of $N=2$ supersymmetry. The expectation 
value of the anomaly is therefore $u$. A more
detailed analysis shows that although one can have several possible
anomalies in the N=2 superfield current
$J$, the only one possible for an
$N=2$ Yang-Mills theory is the chiral one above. 
\par
We noted that this
anomaly is equivalent to the addition of a chiral multiplet which is 
non-reduced i.e. one which does not satisfy the
analogue of equation (3.1). The anomaly $ S$ appears in the current
equation (1.16) in the form
$D^{ij}S\equiv L^{ij}$ and it is straight forward to verify that this
term satisfies the equations $D^{\alpha (i} L^{jk)}=0=
\bar D^{\dot \alpha (i} L^{jk)}$ and so is a complex linear
multiplet.  
\par 
Taking these facts into account equation (3.20) can be written as
$$
\int d^4xd^4\th\, \triangle (A{\del F\over\del A}-2F + cU ) 
+{\rm {c.c}}=0
\eqno(4.2)$$
The chiral superfield $\triangle$ can be shown, using the constraints of
equation (1.4), to be independent of space-time, but dependent on
$\theta$. The  coefficients in the theta expansion are the arbitrary
parameters of the conformal group of equation (1.8). While not all 
the coefficients of $\triangle$ are non-zero, equation (4.2)
implies that the integral over a chiral superspace of  $\triangle$
times a superfield, which is a function of $A$, 
 vanishes for any chiral superfield $A$. As such, one can
conclude that 
$$A {\del F\over\del A}-2F=c U
\eqno(4.3)
$$ 
plus a constant term  and a term linear in $A$. These latter terms
are not consistent with R symmetry and  can also be discarded for
reasons given later.  Hence, if we consider the fields to be
independent of space-time or taking the vacuum expectation value of
the above equation   we find that
$$
a{\del F\over\del a}-2F=c u
\eqno(4.4)
$$
We can determine the constant $c$ by
considering the large field limit
in which perturbation theory is valid. In this regime
$u={1\over 2}a^2$ and after rescaling $F$, as discussed below equation
(3.8), so as to agree with the rest of the literature we have 
$\displaystyle{F={1\over 2} \tau_{cl}a^2+ {i\over 2\pi}
a^2\ln{{a^2}\over\Lambda^2}}$. Hence we recognise that $c=
8\pi i \beta_1$ where $\beta_1$ is the coefficient of  the one-loop
$\beta$-function. 
\par
The above discussion generalises in a rather straightforward 
way to
the case when the N=2 Yang-Mills theory has gauge group 
$G$
spontaneously broken to $U(1)^r$ where $r$ is the rank of $G$. 
The Ward identity of equation (3.6) is the same except 
that the
first  term is now $\bar A_n{\delta \Gamma \over V_n^{ij}}$ where 
$A_n$ and $V_n^{ij},\ n=1,\ldots r$ are the chiral
superfield strength and prepotential  respectively of the $n^{th}$
$U(1)$ factor.  The Ward identity can be manipulated as before with
the  appropriate label and sum over $n$ added to certain terms. 
The expectation value of the 
anomaly is given by $<S>=2 c <Tr W^2>\equiv c
u$ and so we find the constraint  
$$\sum ^r_{m=1}  a_m {\partial F \over \partial  a_m} -2 F 
= 8\pi i \beta_1 u .
\eqno(4.5)$$
\par
It is straightforward to rederive this equation in the presence
 of $N=2$ matter and we hope to do carry this out elsewhere.  Equation
(4.5) is in agreement with references [15]  and [16] where it  was
derived using the hypereliptic curve  and  the Whitham
dynamics associated with the theory, but
 was only established for   the gauge groups
$SU(N)$, $SO(N)$and $Sp(N)$ with certain
$N=2$ matter. 
\par
In this case of a $N=2$ theory with no superconformal anomaly we can
repeat the above steps and then $F$ will obey equation (4.3),
 but with no right hand side as $\beta_1=0$. This equation then 
determines that $ F= d A^2$ where $d$ is a constant. Hence in the case
of the non-anomalous theories equation (4.4) determines the chiral
part of the effective action. This is in agreement with the argument
given in reference [13] that found that the Greens functions of
the chiral sector of these theories are determined by superconformal
invariance up to constants. In the case of N=2 Yang-Mills, one can 
readily show,
by explicitly applying superconformal transformations to the 
Green's functions and demanding that the result vanish that 
 only the  two point Green's function is non-zero. 
\par
The appearance of an elliptic curve in 
 the non-perturbative solution of Seiberg and Witten [9] for $N=2$
Yang-Mills theory prompted the   authors of references [25] to
associate an  integrable system with the non-perturbative solution. In
particular, they identified the solution with Whitham dynamics. 
In reference [26], it was further shown that the Seiberg-Witten
solution was only be identified with Whitham dynamics provided the
function 
$F$ obeyed the equation 
$$ a {\partial F\over \partial a} -2F =
-\sum _n  T_n{\partial F\over \partial T_n}.
\eqno(4.6)$$
where the  $T_n,\ n=0,1\ldots $ are the times of the Whitham dynamics. 
It has been argued [15,16] that one can set $T_n=0,\  n > \ 1$ and, in
the presence of massless $N=2$ matter, $T_0=0$, leaving
only one the term containing
$T_1$  on the right hand side. In the explicit examples studied
[15,16], it has been found, using the explicit form of the
hyperelliptic curve  that this term does indeed equal $ 8\pi i u$ as
required for agreement with equation (4.5). This equation has also
been found [27] to play an important r\^{o}le in the derivation of N=2
Yang-Mills theory from superstring theories. 
\par
The alternative form of the condition of equation (4.6) also has a
natural interpretation in terms of the derivation given in this paper. 
As pointed out at the end of section two we can, in $N=1$
supersymmetric theories, replace the anomaly term by a suitable
derivative with respect to the supergravity  conformal
compensating  chiral  superfield. For the case of $N=2$
supersymmetric theories with the anomaly structure of equation (4.1), 
it is natural  to take the conformal supergravity 
compensator to be  a
non-reduced chiral superfield which we also denoted by $\phi$. The
coupling between the anomaly and the compensator is then  given by
$\kappa \int d^4x d^4\theta \phi S$. We observe that adding this term
to $F$ and  functionally differentiating with respect to $\phi $ and
setting
$\phi=0$ contributes the term $U$ in equation (4.3). In particular,
differentiating with respect to the  highest component, denoted
$t_1$,  of
$\phi$ will result in   $u$. Hence we can replace the anomaly in equation
(4.3) and the term $u$ in equation (4.5) by differentiation with
respect to 
$\phi$ and $t_1$ respectively and  as such, we can cast equation (4.5)
in the form of equation (4.6) if we set $t_1 \propto {\rm {ln}}T_1$.
The above non-reduced  chiral compenstor includes the reduced chiral 
compensator $\phi_r$, discussed at the end of section three that plays
such a special r\^{o}le in the geometry of $N=2$ couplings to
supergravity. Presumeably, the other times $T_n,\ n>1$ that appear in 
 equation (4.6) are related to the components of other
"compensating"   superfields required to construct  the
superstring. 
 We
will return to a more detailed analysis of this point and the
associated supergravity theory in a subsequent paper [28]. 
\par
For the non-conformal theory, equation (4.4) does not determine the
chiral effective action of equation (3.8) completely. In particular,
one is missing the knowledge of $u$ as a function of $A$ which, if
provided, would allow one to solve to write the above differential
equation in terms of only one variable. Very roughly speaking
we are missing about half the information contained in the
solution. Seen from the perspective of reference [26], we are missing
the Whitham dynamics itself. 
It would be interesting to also give a
derivation of this missing information from the basic theory. 
\par
Some papers [16,26] have observed that the constraint of equation
(4.6)  looks similar to the $L_0$ constraint found in matrix models.
It is  thought to be true in matrix models that the entire
solution is given by the imposition of all of the positive Virasoro
constraints, i.e.  
$L_n,\ n \geq 0$. One might also hope that a corresponding statement
is true for the $N=2$ Yang-Mills theory and, in view of the work  of
the current paper, one may wonder if the  constraints 
$L_n,\ n\ >0$  correspond to
Ward Identities for some additional, possibly broken, symmetries of
 the theory. 
\par
It would also be interesting  
to attempt to obtain information about $N=1$ supersymmetric theories
using arguments analogous to those used in section four, but starting
from  equation (2.21). 
\medskip
{\bf {Acknowledgement}}
\medskip
We would like to thank G. Delius, A. Marshakov and K. Takasaki for
discussions and in particular T. Eguchi for drawing to our attention
equation (4.6) and stressing its relavence to the work of this paper. 
\par
Part of this work was carried out while one of the authors (PW)
was  visiting the Yukawa Institute, Kyoto, Japan under the exchange
agreement between Japan Physical Society  of Japan and the Royal
Society  of England and at the Chalmers Institute of Technology,
Goteborg, Sweden funded by the  EU Human Capital and Mobility
Programme no CHRX-CT92-0069.

\medskip
{\bf {References}}
\medskip
\parskip 0pt
\item {[1]} M. Sohnius and P. West Phys. Lett. B100 (1981) 45; 
S.Mandelstam, Nucl. Phys. B213 (1983) 149; P.S. Howe, K.S. Stelle
and P.K. Townsend, Nucl. Phys. B214 (1983) 519; Nucl. Phys. B236
(1984) 125. L. Brink, O. Lindgren and B. Nilsson, Nucl. Phys. B212
(1983) 401.
\item {[2]} P. Howe, K. Stelle and P. West, Phys. Lett 124B
(1983) 55.
\item {[3]} A. Parkes and P. West, Phys. Lett138B (1984) 99,
D. Jones and L.Mezincescu Phys. Lett B138 (1984) 293, D.I.Kazakov,
Phys. Lett B179 (1986) 352,  Mod Phys Lett. A2 (1987) 663, O. Piguet
and K. Sibold, Helv. Phys. Acta 63 (1990) 71. 
\item {[4]} M.T. Grisaru and W.Siegel, Nucl. Phys. B201
(1982) 292, 
 P.S. Howe, K.S. Stelle and P.K. Townsend, 
Nucl. Phys. B214 (1983) 519; Nucl. Phys. B236 (1984) 125,  
 P. West "Supersymmety and Finiteness"  
Proceedings of the
1983 Shelter Island II Conference on Quntum Field Theory and
Fundamental Problems of Physics, edited by R. Jackiw, N. Kuri , S.
Weinberg and E. Witten (M.I.T. Press). 
\item {[5]} C. Montonen and D. Olive, Phys. Lettl 72B (1977) 117.
\item {[6]} H. Osborn, Phys. Lett. B83 (1979) 321, A. Sen, Phys.
Lett. 329B (1994) 217.   
\item {[7]} P. Rossi, Phys. Lett. 99B (1981) 229. 
\item {[8]} P. Argyres, M. Plesser, N. Seiberg and E. Witten,
"New $N=2$ Superconformal Field Theories in Four Dimensions",
hep-th/9511154.
\item {[9]} N. Seiberg and E. Witten, Nucl. Phys. B431 (1994) 484. 
\item {[10]} A. Belavin., A. Polyakov and A. Zamalodchikov, Nucl.Phys.
B241 (1984) 333.
\item {[11]} B. Conlong and P. West, J. Phys. A26 (1993) 3325. 
\item {[12]} P. Howe and P. West, Non-Perturbative Green's
 Functions in Theories with Extended Supersymmetry, hep-th/9509140,
to be published in Int. J. Mod. Phys.; Operator Product Expansions 
in Four-Dimensional Superconformal Field Theories, hep-th/9607060. 
\item {[13]} P. Howe and P. West, Phys. Lett.  B223 (1989) 377. 
\item {[14]} M. Mantone, Phys. Let. 357B (1995) 342.
\item {[15]} J. Sonnenschein, S. Theisen and S. Yankielowicz, 
Phys. Let. 367B (1996) 145. 
\item {[16]} T. Eguchi and S.K. Yang, hep-th/9510138,
\item {[17]} S. Ferrara and B. Zumino, Nucl. Phys. B87 (1975) 2074.
\item {[18]} S. J. Gates, M. Grisaru and W. Seigel and M. Rocek,
Superspace; One thousand and one lessons in superspace, 
Benjamin/Cummings (1983).
\item {[19]} I. Buchbinder and S. Kuzenko, Ideas and Methods of
Supersymmetry and Supergravity, IOP Publishing, 1995;  
O. Piguet and K. Sibold, "Renormalized Supersymmetry" in Progress 
in Physics, vol 12, Birkhauser Boston Inc, 1986. 
\item {[20]} A. Salam and J. Strathdee, Phys. Lett. 51B (1974) 353;
P. Fayet, Nucl. Phys. B113 (1976) 135.
\item {[21]} L. Mezincescu, JINR preprint P2 (1979) 1257, 
\item {[22]} 
P. Howe, K. Stelle and P. Townsend, Nucl. Phys. B236 (1984) 125; 
J.Koller, Nucl. Phys. B222 (1983) 319, 
Phys. Lett. 124B (1983) 324.  
\item {[23]} A. Galperin, E. Ivanov, S. Kalitzin, V. Ogievetsky
and E. Sokatchev, Class. Quant. Grav. 2 (1985) 155,  Class. Quant.
Grav.1 (1984) 469.
\item {[24]} N. Seiberg, Electromagnetic Duality in Supersymmetric
Non-abelian Gauge Theories, hep-th/9411149
\item {[25]} A. Gorsky, I Krichiver, A. Marshakov and A. Morozov,
Phys. Lett. B355 (1995) 466.  
\item {[26]} T. Nakatsu and K. takasaki, Whitham-Toda Hierachy
and $N=2$ Supersymmetric Yang-Mills Theory, hep-th/9509162. 
\item {[27]} S. Kachru, A. Klemm, W. Lerche, P. Mayr and C. Vafa,
Non-perturbative Results on the Point Particle Limit of $N=2$
Heterotic String Compactification, hep-th/9507144. 
\item {[28]} P. Howe and P. West, in preparation.
\item {[29]} A. Van Proeyen, 
"Supersymmetry and Supergravity 1983", ed
B. Milewski, World Scientific, 1983. 
\item {[30]} P. West, "Introduction to Supersymmetry and
Supergravity", World Scientific, 1986.

\end